\documentclass[
superscriptaddress,secnumarabic,
amssymb,amsmath,nobibnotes,aps,prd,showkeys,showpacs,nofootinbib]{revtex4}%
\usepackage{graphicx}
\usepackage{epsf}
\usepackage{bm}
\usepackage{amsmath}
\usepackage{amsfonts}
\usepackage{amssymb}
\usepackage{epstopdf}
\usepackage{natbib}
\usepackage{color}
\usepackage{float}
\providecommand{\U}[1]{\protect\rule{.1in}{.1in}}

\newcommand{\be}{\begin{equation}}
\newcommand{\ee}{\end{equation}}

\newcommand{\mincir}{\raise
-3.truept\hbox{\rlap{\hbox{$\sim$}}\raise4.truept\hbox{$<$}\ }}
\newcommand{\magcir}{\raise
-3.truept\hbox{\rlap{\hbox{$\sim$}}\raise4.truept\hbox{$>$}\ }}

\begin{document}

\title{Note on the reheating temperature in Starobinsky-type potentials}

\author{Jaume   Haro}
\email{jaime.haro@upc.edu}
\affiliation{Departament de Matem\`atiques, Universitat Polit\`ecnica de Catalunya, Diagonal 647, 08028 Barcelona, Spain}

\author{Llibert Arest\'e Sal\'o}
\email{l.arestesalo@qmul.ac.uk} 
\affiliation{School of Mathematical Sciences, Queen Mary University of London, Mile End Road, London, E1 4NS, United Kingdom}

\thispagestyle{empty}

\begin{abstract}
The relation between the reheating temperature, the number of e-folds and the spectral index is shown for the Starobinsky model and some of its descendants through a very detailed calculation of these three quantities. The conclusion is that for viable temperatures between $1$ MeV and $10^9$ GeV the corresponding values of the spectral index enter perfectly in its $2\sigma$ C.L., which shows the viability of this kind of models.

\end{abstract}

\vspace{0.5cm}

\maketitle
\section{Introduction}

{The Starobinsky model based on $R^2$-gravity in the Jordan frame \cite{Starobinsky}, which was extensively studied in the literature (see for instance \cite{felice, vilenkin,nojiri,nojiri1}
and \cite{aho} for a detailed dynamical analysis),  is one of the most promising scenarios to explain the inflationary paradigm  proposed by A. Guth in \cite{guth} because it provides theoretical data about the power spectrum of perturbations, which matches very well with the recent observational data obtained by the Planck team \cite{Planck}.
In addition, contrary to the Guth's paper, in \cite{Starobinsky} the author briefly details a successfully reheating mechanism based on the production of particles named {\it scalarons} whose decay products reheat the universe (see \cite{Starobinsky1,vilenkin,haro} for a detailed discussion of this mechanism),
obtaining a reheating temperature around $10^9$ GeV \cite{Gorbunov} (see also \cite{felice} for the derivation of this reheating temperature when the decay products are massless and  minimally coupled with gravity).

\

Working in the Einstein frame, $R^2$-gravity leads to the well-known {\it Starobinsky potential} \cite{felice}, which has been recently studied as an inflationary potential, and the reheating temperature provided by the model is related to its corresponding spectral index \cite{german, cook,rg} (see also \cite{oku} for the calculation of the reheating temperature when inflation come from a 
 constant-roll  era). However,  contrary to  \cite{hashiba,kolb2} where the authors consider the gravitational production of superheavy particles, in those papers the reheating mechanism is not taken into account; instead of it, it is assumed that during the oscillations of the inflaton field the effective Equation of State (EoS) parameter is constant. 
From our viewpoint, it is difficult to 
understand how it is possible to make any meaningful statements about
reheating temperature without consideration of its concrete mechanisms,
apart from the hypothesis of instant thermalization \cite{evan}, which
has to be still justified \cite{Starobinsky2}.

\

Anyway, although we do not discuss any reheating mechanism, the main goal of this note is to review these papers and find a very precise relation between the reheating temperature and the number of e-folds as a function of the spectral index of scalar perturbations, especially for the Starobinsky-type potentials that we have proposed by slightly modifying the Starobinsky potential so that its behavior near the origin is as a power law potential.

}

\

The work is organized as follows: In Section II we perform a very accurate calculation of the number of e-folds from the moment in which the pivot scale leaves the Hubble horizon to the end of inflation, which will be used in Section III to relate the spectral index provided by the Starobinsky-type potentials with its reheating temperature. And we show numerically that for temperatures between $1$ MeV and $10^9$ GeV the spectral index ranges in its $2\sigma$ Confidence Level, which means that these reheating temperatures are compatible with the model. Section IV is devoted to the study of the particular case where the effective EoS parameter during the oscillations of the inflaton field is equal to $1/3$. This is a very particular case where it is impossible to define exactly when the radiation starts and, thus,  that it is impossible to obtain the value of the reheating temperature. From what we show, we might argue that this case is physically unacceptable and all its consequences derived from it must be disregarded. However, one has to take into account that a constant effective EoS during the oscillations of the inflaton is only an approximation because the physics of this period is far from being clearly understood and, thus, this approximation could lead to wrong conclusions.
 Finally, in the last section we discuss the obtained results.

\

The units used throughout the paper are $\hbar=c=1$ and   the reduced Planck's mass  is denoted by 
$M_{pl}\equiv \frac{1}{\sqrt{8\pi G}}\cong 2.44\times 10^{18}$ GeV.

\

{

\section{The number of e-folds}
First of all, we will assume that from the end of inflation to the beginning of the radiation era the effective Equation of State (EoS) parameter, namely $w_{re}$ following the notation of \cite{cook}, is constant. However,
from the end of inflation to the onset of the radiation era
 there is a transient period
where the EoS is not constant. This period is largely unknown, as well as the mechanisms to produce and thermalize the relativistic plasma which reheats the universe. So, taking $w_{re}$ constant is an approximation which in some cases could lead to incorrect results and interpretations.

In this situation,
when $w_{re}\not=1/3$, the number of e-folds from the moment in which the pivot scale crosses the Hubble horizon to the end of inflation, namely $N_k$, 
is given by (see formula (2.4) of \cite{rg})
\begin{eqnarray}
N_k=\ln(a_{eq}/a_k)+\frac{\ln(\rho_{eq}/\rho_{end})}{{3(1+w_{re})}}+ \frac{3w_{re}-1}{12(w_{re}+1)}\ln(\rho_{eq}/\rho_{re}),
\end{eqnarray}
where ``eq" means the matter-radiation equality and ``end" the end of the inflationary period (see also \cite{dai,munoz}).

This expression could be written as 
\begin{eqnarray}
N_k=-\ln(1+z_{eq})+\ln(H_k/k_{phys}) +\frac{\ln(\rho_{eq}/\rho_{end})}{{3(1+w_{re})}}
+ \frac{3w_{re}-1}{12(w_{re}+1)}\ln(\rho_{eq}/\rho_{re}),
\end{eqnarray}
where $z$ denotes the red-shift and $k_{phy}$ is the physical value of the pivot scale.

\,

We choose for example $k_{phys}\equiv \frac{k}{a_0}=0.05 \mbox{Mpc}^{-1}\cong 1.31\times 10^{-58} M_{pl}$, $z_{eq}=3365$, 
$\rho_{eq}=\frac{\pi^2}{15}g_{eq}T_{eq}^4$ with $g_{eq}=3.36$ and, 
from the adiabatic evolution of the universe  after reheating, we  have that $a_{eq}T_{eq}=a_0T_0\Longrightarrow T_{eq}=(1+z_{eq})T_0$, where the present CMB temperature is 
$T_0=2.725 \mbox{ K}\cong 2.35\times 10^{-4}$ eV.

\

We also consider $\rho_{re}=\rho_{\gamma,re}$ where $\rho_{\gamma,re}=\frac{\pi^2}{30}g_{re}T_{re}^4$ is the energy density of the relativistic plasma (see for example the formula (3.51) of Mukhanov's book  \cite{mukhanov}) at the reheating time
and $g_{re}=g(T_{re})$ is the effective number of degrees of freedom at the beginning of the radiation epoch. This is verified since after inflation the inflaton field has completely decayed and, thus, $\rho_{\phi}$ plays no roll.


\

We use as well that $\rho_{end}=\frac{3}{2}V_{end}$ and, in order to get the value of $H_k$, we need the spectrum of scalar perturbations when the pivot scale crosses the Hubble horizon \cite{btw}, namely
\begin{eqnarray}\label{power}
\mathcal{P}_{\zeta}=\frac{H_k^2}{8\pi^2M_{pl}^2\epsilon_k}\cong
 2\times 10^{-9},
\end{eqnarray}
where 
\begin{eqnarray}
\epsilon_k=\frac{M_{pl}^2}{2}\left( \frac{V_{\phi}(\phi_k)}{V(\phi_k)} \right)^2,
\end{eqnarray}
is the main slow-roll parameter at the crossing time.

\

Then, we can write the number of e-folds as follows:
\begin{eqnarray} \label{Nkw}
N_k=-\ln(1+z_{eq})+\ln(H_k/k_{phys}) +\frac{1}{4}\ln(\rho_{eq}/\mbox{GeV}^4)
+\frac{\ln(\mbox{GeV}^4/\rho_{end})}{{3(1+w_{re})}}
+ \frac{3w_{re}-1}{12(w_{re}+1)}\ln(\mbox{GeV}^4/\rho_{re})\nonumber\\
\cong 96.5684+\frac{1}{2}\ln \epsilon_k
+\frac{\ln(\mbox{GeV}^4/\rho_{end})}{{3(1+w_{re})}}
+ \frac{3w_{re}-1}{12(w_{re}+1)}\ln(\mbox{GeV}^4/\rho_{re}),
\end{eqnarray}
which only depends on the main slow roll parameter when the pivot scale leaves the Hubble horizon, the effective EoS parameter  $w_{re}$, the energy density at the end of inflation and the reheating temperature.

\

\section{Different models}

We will consider the following kind of Starobinsky-like potentials, depicted in Figure \ref{fig:V}, 
\begin{eqnarray}\label{starobinsky}
V_n(\phi)=\lambda_n M_{pl}^4
(1-e^{-\kappa_n \phi^n/M_{pl}^n})^2,
\end{eqnarray}
where $\lambda_n$ and $\kappa_n$ are  dimensionless parameters (see also \cite{sebastiani} for the study of other potentials  slightly different  from the Starobinsky one). As we have pointed out in the introduction, these potentials represent a variation of the Starobinsky potential (the one when $n=1$) with regards to the power of the scalar field. This $n$ parameter enables us to mimic the behavior of a power law potential near the origin. Note that the factor $\kappa_1$ is required to be $\sqrt{2/3}$ in the Starobinsky model in order to impose canonical normalization of the scalar $\phi$ when passing from the $R^2$ theory to the Einstein frame  \cite{felice}. For $n\neq 1$, given that this is not the case, we will be considering different possible factors so as to discuss for which ones the observations constraints are best fulfilled.

\

,
\begin{figure}[H]
\begin{center}
\includegraphics[scale=0.6]{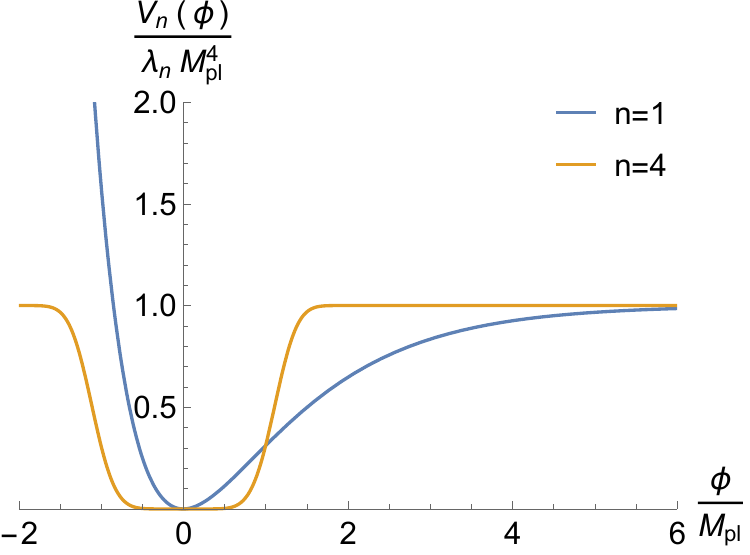}
\end{center}
\caption{Different shapes of the potential for even and odd values of $n$, here considering $\kappa_4=\kappa_1$. 
}
\label{fig:V}
\end{figure}

In Figure \ref{fig:dynamics} we see that for $n$ even (the odd case is clear) the inflaton field oscillates in the deep well potential after inflation, thus leaving its energy in order to produce enough particles to reheat the universe.

\begin{figure}[H]
\begin{center}
\includegraphics[scale=0.6]{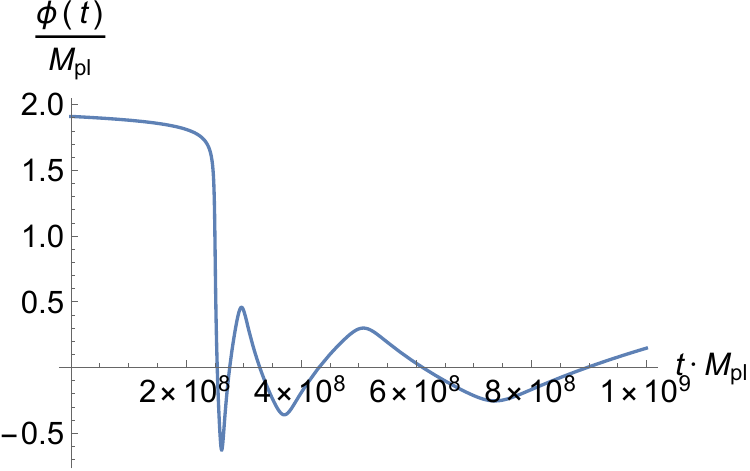}
\end{center}
\caption{The dynamical evolution of the inflaton field for $n=4$.}
\label{fig:dynamics}
\end{figure}

\

This kind of potentials, contrary to the power law ones,  are allowed by the observational  Planck results because the values of the spectral index  $n_s$ and the ratio of tensor to scalar perturbations $r$ enter perfectly  in the 
marginalized confidence contour in the plane $(n_s, r)$ at $1\sigma$ and $2 \sigma$ Confidence Level. 

\

In addition, near the origin the potential is like $\phi^{2n}$, that is, the shape of the well of the Starobinsky-type potential is the same as for a power law potential. Then, during the oscillations of the inflaton field, for a a potential $V(\phi)=V_0\phi^{2n}$ and using the virial theorem, we get that the effective EoS parameter is given by
\cite{turner, ford} 
\begin{eqnarray}
w_{re}=\frac{n-1}{n+1},
\end{eqnarray}
meaning that this also holds for the potentials (\ref{starobinsky}).

\

On the other hand, dealing with the power spectrum of scalar perturbations, 
we have that
\begin{eqnarray}\label{epsk}
\epsilon_k= 2(\kappa_nn)^2\left( \frac{\phi_k}{M_{pl}} \right)^{2(n-1)}\frac{e^{-2\kappa_n \phi^n_k/M_{pl}^n}}{\left(1-e^{-\kappa_n \phi^n_k/M_{pl}^n}\right)^2 }
\cong 2(\kappa_nn)^2\left( \frac{\phi_k}{M_{pl}} \right)^{2(n-1)}e^{-2\kappa_n\phi^n_k/M_{pl}^n},\end{eqnarray}
and 
\begin{eqnarray}
\eta_k=M_{pl}^2\frac{V_{\phi\phi}(\phi_k)}{V(\phi_k)}=2\kappa_nn(n-1)\left( \frac{\phi_k}{M_{pl}} \right)^{n-2}\frac{e^{-\kappa_n \phi^n_k/M_{pl}^n}}{1-e^{-\kappa_n \phi^n_k/M_{pl}^n} }
\\-
2(\kappa_nn)^2
\left( \frac{\phi_k}{M_{pl}} \right)^{2(n-1)}e^{-\kappa_n \phi^n_k/M_{pl}^n}\frac{1-2e^{-\kappa_n \phi^n_k/M_{pl}^n}}{\left(1-e^{-\sqrt{\frac{2}{3}} \phi^n_k/M_{pl}^n}\right)^2 }
\cong -2(\kappa_nn)^2\left( \frac{\phi_k}{M_{pl}} \right)^{2(n-1)}e^{-\kappa_n\phi^n_k/M_{pl}^n},\end{eqnarray}
and, thus, the spectral index can be computed for $n=1$ without the approximation carried out in the last step for both $\epsilon_k$ and $\eta_k$. Hence, using the well known relation at first order  between  
the spectral index and these slow roll parameters $n_s=1+2\eta_k-6\epsilon_k$ (see for example \cite{btw}), we obtain that
\begin{eqnarray}\label{phik}
1-n_s\cong 6\epsilon_k-2\eta_k=
\frac{8}{3}\frac{e^{-\sqrt{\frac{2}{3}} \phi_k/M_{pl}}}{1-e^{-\sqrt{\frac{2}{3}} \phi_k/M_{pl}}}\left(
\frac{e^{-\sqrt{\frac{2}{3}} \phi_k/M_{pl}}}{1-e^{-\sqrt{\frac{2}{3}} \phi_k/M_{pl}}}+1
\right), 
\end{eqnarray}
getting that $\phi_k=\sqrt{\frac{3}{2}}\ln\left(\frac{7-3n_s+4\sqrt{4-3n_s}}{3(1-n_s)} \right)$ (see for instance \cite{german}). Effectively, let us express $\epsilon_k=\frac{4}{3(1-s)^2}$ and $\eta_k=\frac{4}{3}\frac{2-s}{(1-s)^2}$, where $s=e^{\sqrt{\frac{2}{3}}\frac{\phi_k}{M_{pl}}}$. So, equation \eqref{phik} can be written as a 2nd order polynomical equation, namely $3 (n_s-1)s^2+s(-6n_s+14)+3n_s+5=0$, which is satisfied for $s=\frac{7-3n_s+4\sqrt{4-3n_s}}{3(1-n_s)}$, from which the given value of $\phi_k$ follows.

On the other hand,  for $n\not=1$ one approximately has
\begin{eqnarray}\label{phikn}
1-n_s\cong 6\epsilon_k-2\eta_k\cong 4(\kappa_nn)^2\left( \frac{\phi_k}{M_{pl}} \right)^{2(n-1)}e^{-\kappa_n \phi^n_k/M_{pl}^n}.
\end{eqnarray}

Only for the  exact Starobinsky model ($n=1$) one can express analytically  $\epsilon_k$ as a function of $1-n_s$. In the other cases ($n\not= 1$) one has to obtain it numerically.

\

Note also that inflation ends when
\begin{eqnarray}\label{end}
\epsilon_{end}=2(\kappa_nn)^2\left( \frac{\phi_{end}}{M_{pl}} \right)^{2(n-1)}\frac{e^{-2\kappa_n \phi^n_{end}/M_{pl}^n}}
{\left(1-  e^{-\kappa_n\phi^n_{end}/M_{pl}^n} \right)^2}=1
\end{eqnarray}
and the value of $\phi_{end}$ can only be obtained analytically for the exact Starobinsky model.

\subsection{Case $n=1$: The exact Starobinsky model}

  As we have already explained in the introduction, this potential comes from $R^2$-gravity in the Einstein frame (see for example \cite{riotto} for a detailed explanation) and, 
since $n=1$, $ w_{re}=0$.  In addition, from (\ref{end}) one gets
\begin{eqnarray}
\phi_{end}=-\sqrt{\frac{3}{2}}\ln(\sqrt{3}(2-\sqrt{3}))M_{pl}\cong 0.9402 M_{pl},
\end{eqnarray}
obtaining 
\begin{eqnarray}
V_{end}=4\lambda (2-\sqrt{3})^2M_{pl}^4\Longrightarrow \rho_{end}=6\lambda (2-\sqrt{3})^2 M_{pl}^4,\end{eqnarray}
where we have used that at the end of inflation $\dot{\phi}_{end}^2=V(\phi_{end})$ and, thus, $\rho_{end}=\frac{3}{2}V(\varphi_{end})$.

\

To calculate  the value of the parameter $\lambda$ we use that $H_k^2\cong \frac{V(\phi_k)}{3M_{pl}^2}\cong \frac{\lambda{M_{pl}^2}}{3}$. Therefore, from the formula of the power spectrum of scalar perturbations (\ref{power}) we obtain
\begin{eqnarray}
\lambda\cong 48 \pi^2\times 10^{-9}\epsilon_k,
\end{eqnarray}
where $\epsilon_k=\frac{4}{3(1-s)^2}$, being $s=\frac{7-3n_s+4\sqrt{4-3n_s}}{3(1-n_s)}$, as used in \eqref{phik}. With regards to the number of e-folds, it is given by
\begin{eqnarray}\label{e-foldsstarobinsky}
N_k\cong 96.5684+\frac{1}{2}\ln\epsilon_k+\frac{1}{3}\ln\left(  \frac{\mbox{GeV}^3\rho_{re}^{1/4}}{\rho_{end}}  \right),
\end{eqnarray}
but can also be calculated using the formula
\begin{eqnarray} \label{Nkeps}
N_k=\int_ {t_k}^{t_{end}}Hdt=\frac{1}{M_{pl}}\int_{\phi_{end}}^{\phi_{k}}\frac{1}{\sqrt{2\epsilon}}d\phi.
\end{eqnarray}
So, using the values defined above, one gets that
\begin{eqnarray}\label{e-foldsstarobinsky1}
N_k = \frac{1}{4}\left(3\left(e^{\sqrt{\frac{2}{3}}\phi_k}-e^{\sqrt{\frac{2}{3}}\phi_e}\right)-\sqrt{6}(\phi_k-\phi_e) \right),
\end{eqnarray}
which leads to $44.02\leq N_k\leq 54.88$ for the values of $n_s$ given by Planck's team \cite{planck18} within its $2\sigma$ C.L., namely $0.9565\leq n_s\leq 0.9733$. We note that if we invert this function the obtained result coincides to a great extent with the relation in equation $(32)$ of \cite{roest}, reached through a next-to-leading order expansion.

\

Now, by equating (\ref{e-foldsstarobinsky}) and (\ref{e-foldsstarobinsky1}) we get a relation between the reheating temperature and the spectral index of scalar perturbations, which is represented in Figure \ref{fig:temperature}.

\

Here, it is important to take into account that a lower bound of the reheating temperature is $1$ MeV because the Big Bang Nucleosynthesis (BBN) occurs at this scale and the universe needs to be reheated at this epoch. In the same way the upper bound of the reheating temperature could be obtained imposing that relic products such as gravitinos or modulus fields  which  appear in supergravity or string theories do not affect the BBN success, which happens for reheating temperatures below $10^6$ TeV (see for instance \cite{gkr}).

\begin{figure}[H]
\begin{center}
\includegraphics[scale=0.8]{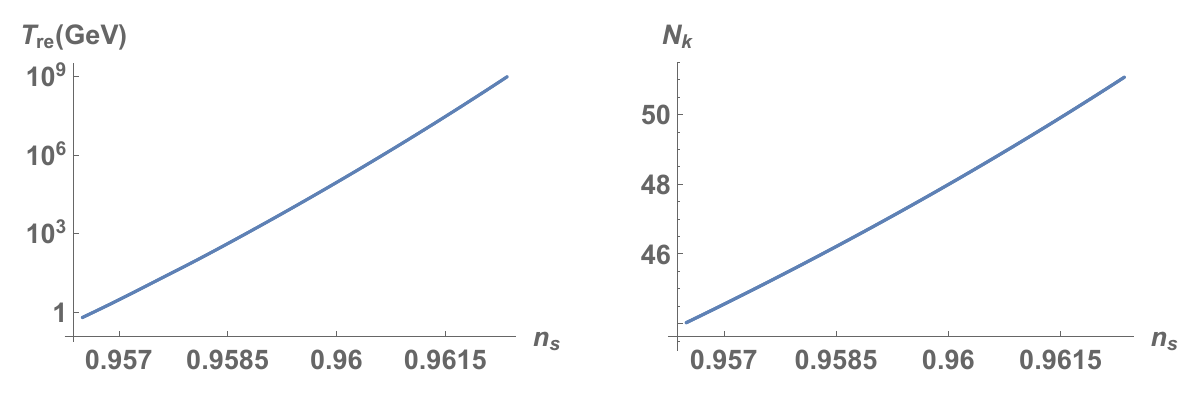}
\end{center}
\caption{The reheating temperature and the number of e-folds for $n=1$ as a function of the spectral index, only for temperatures between $1$ MeV and $10^6$ TeV.}
\label{fig:temperature}
\end{figure}

\

In Figure \ref{fig:temperature} we can see that for reheating temperatures between $1$ MeV and $10^6$ TeV the spectral index satisfies
$0.9565<n_s<0.9624$, which enters perfectly in its $2\sigma$ C.L., and the number of e-folds ranges between $44.02$ and $51.08$, which is in agreement 
 with the previous and maybe not so exact calculations made in \cite{cook,rg} and coincides as well to a great extent with the result obtained in \cite{german} by using a diagrammatic approach (see the reheating temperature shown in Figure $3$ of \cite{german}).
Note also that we have used as $g_{re}$ the function obtained as a linear interpolation of the values in the Table $1$ of \cite{husdal}.

\

{ We end this subsection pointing out that the Starobinsky potential could also be used in quintessential inflation improving the well-known Peebles-Vilenkin model
\cite{pv}. In that case it was shown in \cite{ha} that the reheating temperature depends on the mechanism used to reheat the universe. More precisely, when superheavy particles (whose decay products will reheat the universe) are gravitationally produced, the upper bound of $T_{re}$ is  around $40$ TeV and, when the mechanism is the so-called {\it instant preheating} \cite{fkl0,fkl}, one gets the following lower bound, $T_{re}\geq 20$ TeV.}

\subsection{Case $n\not= 1$}

When $n\not=1$ the relation between the reheating temperature and the spectral index has to be calculated numerically. For each value of $n_s$ in 
the $2\sigma$ C.L. interval, we have numerically solved equations \eqref{phik} and \eqref{end} in order to find the values of $\phi_k$ and $\phi_{end}$. Then we have used the value of $\epsilon_k$ in equation \eqref{epsk} in order to calculate the number of e-folds $N_k$ as stated in (16). And finally we have obtained the reheating temperature by setting this value equal to the one in equation \eqref{Nkw}. As in the case $n=1$ we have taken as $g_{re}$ the linear interpolations of the values in the table $1$ of \cite{husdal}.

\

In Figure \ref{fig:TrNkns}, taking viable reheating temperatures from $1$ MeV to $10^6$ TeV, we have depicted the corresponding values of the spectral index for several models and several values of $\kappa_n$, showing that they enter in its $2\sigma$ C.L. We have also represented the corresponding number of e-folds for these values of $n_s$. The models studied correspond to the values $n=3,4$ and $5$ which are respectively equal to the following values of the effective EoS parameter, $w_{re}=1/2$, $3/5$ and $2/3$. Note that in all these cases the reheating temperature decreases as $n_s$ grows, in opposite to what happens when $n=1$. This arises from the fact that the last term in equation \eqref{Nkw} vanishes for $n=2$. As a consequence, $T_{re}$ is constant in $n_s$ for $n=2$, thus increasing (resp. decreasing) as a function of $n_s$ for $n<2$ (resp. $n>2$).

\

For each of these values of $n$ we have studied the results for values of $\kappa_n$ between $0.2$ and $10$ and we have also drawn straight lines for the bounds for the reheating temperature as well as the lower limit of the allowed interval for $n_s$ at $1\sigma$ C.L. according to the results of \cite{planck18}, given that all the depicted values of $n_s$ are already in the $2\sigma$ C.L. interval. While for all the values of $n$ and $\kappa_n$ that we have represented the allowed values of the reheating temperatures fall within the $2\sigma$ C.L. interval of $n_s$, when both $n$ and $\kappa_n$ become higher a wider range of the allowed reheating temperatures correspond to a value of $n_s$ within the $1\sigma$ C.L. interval. With regards to the number of efolds, all the obtained values (namely between $55$ and $65$) are feasible. And, as far as the ratio of the tensor to scalar perturbations is concerned, it does not influence our results since in all the cases it is verified that $r<10^{-5}$.

\begin{figure}
\begin{center}
\textbf{$n=3$}\par\medskip
\includegraphics[scale=0.43]{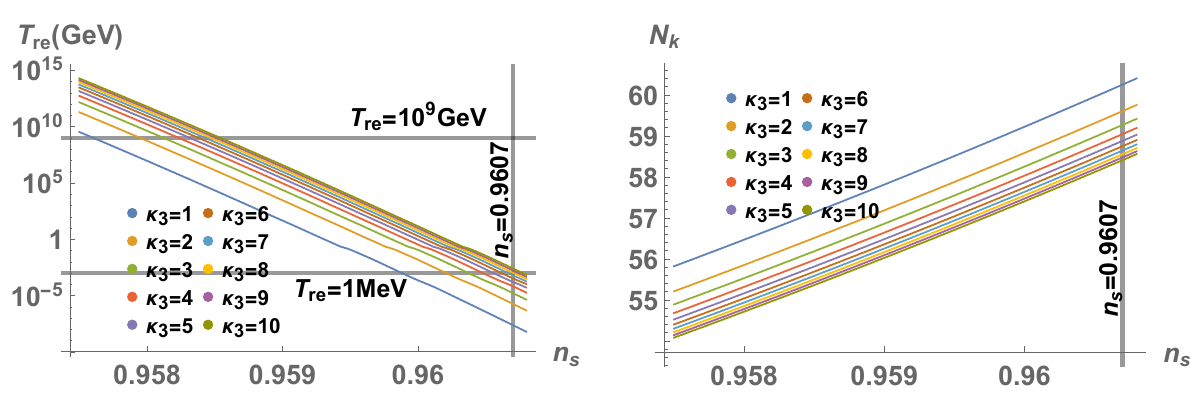}
\includegraphics[scale=0.43]{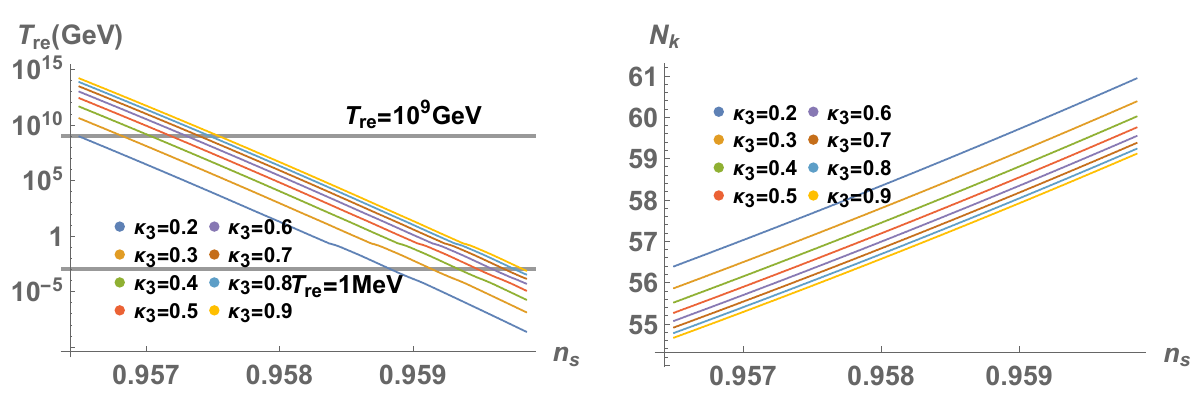}
\textbf{$n=4$}\par\medskip
\includegraphics[scale=0.43]{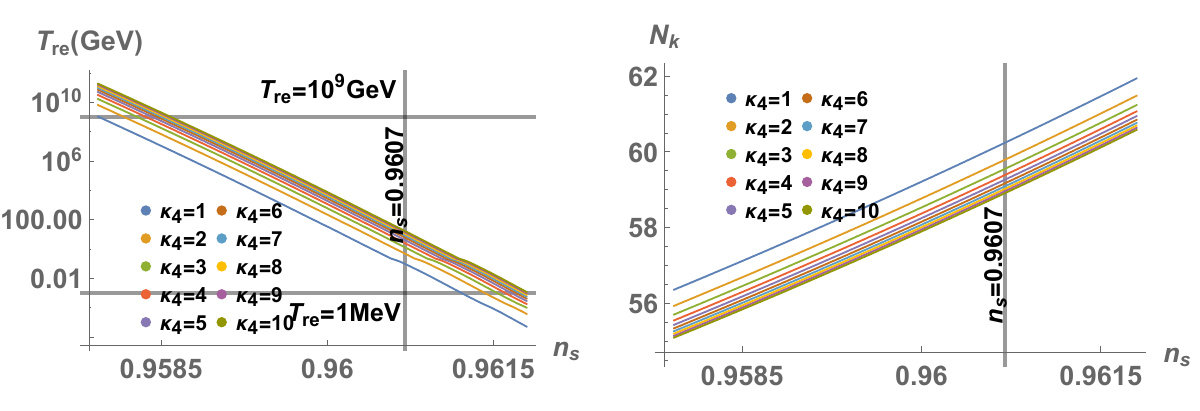}
\includegraphics[scale=0.43]{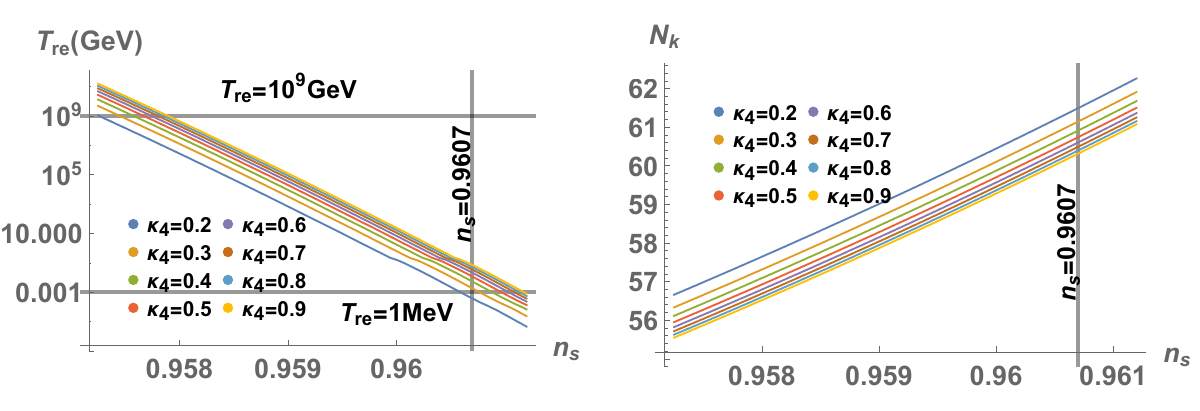}
\textbf{$n=5$}\par\medskip
\includegraphics[scale=0.43]{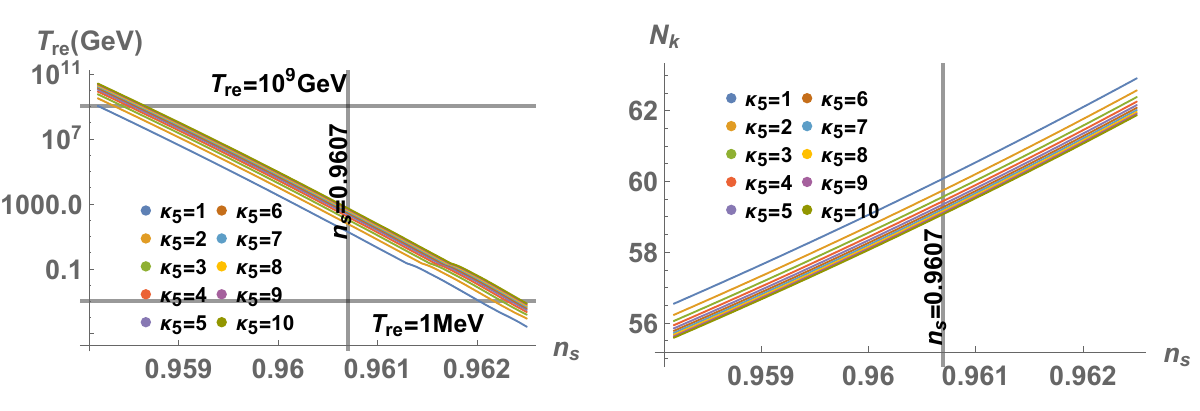}
\includegraphics[scale=0.43]{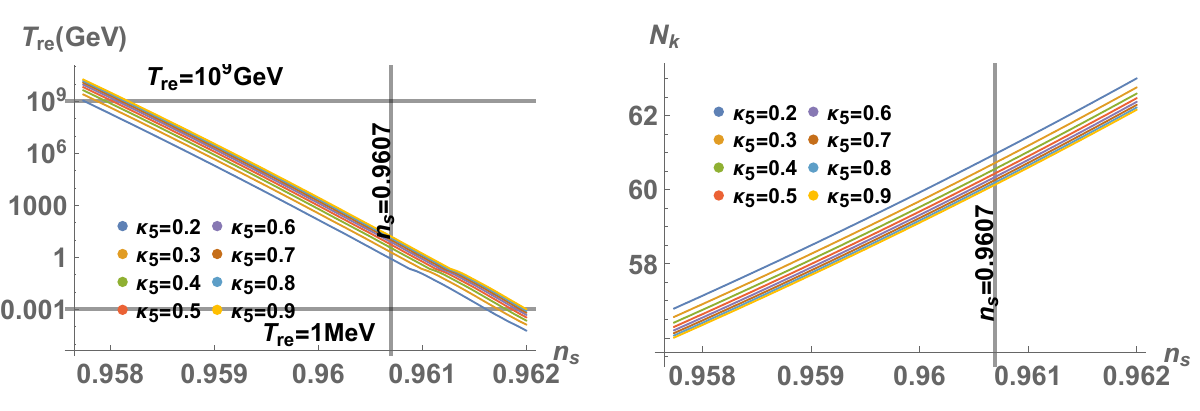}
\end{center}
\caption{The reheating temperature and the number of e-folds as a function of the spectral index, for $n=3$, $4$ and $5$ for temperatures between $1$ MeV and $10^6$ TeV and values of $\kappa_n$ between $0.2$ and $10$. 
Here we have used 
$0.9565\leq n_s\leq 0.9733$, which corresponds to the $2\sigma$ C.L. interval of Planck 2018 data
 \cite{planck18}.}
\label{fig:TrNkns}
\end{figure}

\

Therefore, we see that by modifying the power law behavior at the origin of the Starobinsky potential, we obtain values of the reheating temperature and the number of e-folds from the crossing of the Hubble horizon of the pivot scale until the end of inflation which continue being in accordance with the allowed ones by taking the spectral index within the $2\sigma$ CL of the Planck 2018 data \cite{planck18}. So, we have found a new group of potentials which match as well as the Starobinsky potential with the observational data and, moreover, they contain a parameter $n$ which can be tunned in order to adjust the behavior that we want to have near the origin.

\section{ The particular case $w_{re}=1/3$}

This situation is obtained for our potentials when $n=2$ and it has been already shown that it is impossible to obtain neither the value of the reheating temperature $T_{re}$, nor
the number of e-folds from the end of inflation to the beginning of the radiation era $N_{re}=\ln(\frac{a_{re}}{a_{end}})$. The reason is that, in order  to obtain the values of $T_{re}$ and 
$N_{re}$, one needs to know  the  beginning of the radiation epoch, i.e., when the energy density of the light particles obtained from the decay of the inflaton  field starts to dominate, which does not happen in this case because during the oscillations of inflaton the effective EoS parameter is the same as in the radiation era
\cite{rg,cook,german}.

\

However, in this particular case it is possible to calculate the effective number of degrees of freedom at the beginning of reheating, which is obtained using the formula
 (2.12) of \cite{cook}:
\begin{eqnarray}\label{g}
g_{re}=\left(\frac{43}{11}\right)^4\left(\frac{\pi^2}{30}\right)^{3}\left(\frac{H_ka_0T_0}{e^{N_k} \rho^{1/4}_{end}k } \right)^{12}.
\end{eqnarray}

\

Now, taking into account that $H_k/k=1/a_k$ and that $a_ke^{N_k}=a_{end}$, one gets 
\begin{eqnarray}
g_{re}=\left(\frac{43}{11}\right)^4\left(\frac{\pi^2}{30}\right)^{3}\left(\frac{a_0T_0}{a_{end} \rho^{1/4}_{end} } \right)^{12}
\end{eqnarray} 
and, using that from the end of inflation to the matter-radiation equality the effective EoS parameter is $1/3$, which implies 
$a_{end} \rho^{1/4}_{end}=a_{eq} \rho^{1/4}_{eq}$, one finally obtains
\begin{eqnarray}
g_{re}=\left(\frac{43}{11}\right)^4\left(\frac{\pi^2}{30}\right)^{3}\left((1+z_{eq})\frac{T_0}{\rho_{eq}^{1/4}} \right)^{12}.
\end{eqnarray}

\

This formula is very interesting because it depends neither on the shape of the potential during inflation, nor on the pivot scale. Instead it only depends 
on the number of degrees of freedom at the matter-radiation equality.  Effectively, using once again that $T_{eq}=(1+z_{eq})T_0$ and $\rho_{eq}=\frac{\pi^2}{15}g_{eq}T_{eq}^4$, with $g_{eq}=3.36$ the number of degrees of freedom at the matter-radiation equality,  we get the following abnormally small number
\begin{eqnarray}
g_{re}=\frac{43}{11}\left(\frac{43}{22 g_{eq}}  \right)^3\cong 0.6256,
\end{eqnarray}

which is in contradiction with the values of the effective degrees of freedom (see for instance Figure $1$ of \cite{husdal}). In fact its minimum value is approximately $g_{eq}=3.36$, which is obtained at the matter-radiation equality.

\

Therefore, one might conclude  that the case $w_{re}=1/3$ has to be disregarded, as well as all its consequences. For example, 
the assumption that the value of $g_{re}$ is approximately $100$ (see for instance the Section 2.1 of \cite{cook}) and also the consequences derived in Section 
V of \cite{german}.
However, as we have already explained at the end of the Introduction and at the beginning of Section 2, one has to be cautious with this kind of result because a constant effective EoS is only an approximation. Hence, in order to be sure of their viability, one must  deal with a more  realistic model containing  a well defined reheating mechanism telling us which is the real evolution of the effective EoS parameter from the end of inflation to the beginning of the radiation era (see for example \cite{Lazanov} where the authors study some viable models obtaining numerically the evolution of the effective EoS parameter during this period) .

\

\

}

\section{Conclusions}

In this  short note we have proved that for Starobinsky-type potentials 
of the form $\lambda_n M_{pl}^4(1-e^{-\kappa_n \phi^n/M_{pl}^n})^2$ depending on two  dimensionless parameters $\lambda_n$ and  $\kappa_n$ (which seem to be the best for predicting the values of the power spectrum of perturbations according to the recent observations) the reheating temperature ranges 
in a wide region of its allowed values, which span below $10^6$ TeV -in order that the production of relics such as gravitinos or modulus fields in supergravity theories do not affect the success of the BBN- and above $1$ MeV to ensure that the reheating was previous to the BBN.
In fact, as one can see from Figure 4, the higher the values of $n$ and $\kappa_n$ are, a wider range of allowed reheating temperatures enters in the $1\sigma$ C.L. of the spectral index, indicating that in this sense the model is more favored. 
In addition, for the special case $n=1$ (the Starobinsky model) our results are in agreement with the ones obtained independently in \cite{german} by following a different scheme  named diagrammatic approach.

 \
 
Finally, we have also studied the particular case when the effective EoS parameter during the oscillations of the inflaton field is equal to $1/3$ showing that this case leads to an absurd value of the number of degrees of freedom at the reheating time, meaning that this ideal situation (in a more realistic model the effective EoS is not constant) and its consequences must be disregarded.

 \

\

{\it Acknowledgments.}   We would like to thank Professor Starobinsky for carefully reading our manuscript and also for his comments and suggestions that have been very helpful for improving our work,  to Gabriel Germ\'an for useful conversations, and also to the referee for its criticism which has been very important to improve our work.
 This investigation has been supported by MINECO (Spain) grant
 MTM2017-84214-C2-1-P and  in part by the Catalan Government 2017-SGR-247.

\


\begin{thebibliography}{99}

\bibitem{Starobinsky}
A. A. Starobinsky, 
{\it A new type of isotropic cosmological models without singularity}, 
Phys. Lett. {\bf B91}, 99 (1980).


\bibitem{felice}
A. De Felice and S. Tsujikawa,
{\it f(R) theories},
Living Rev. Rel. {\bf 13},  3  (2010)
	[arXiv:1002.4928 [gr-qc]].
	
\bibitem{vilenkin}	
A. Vilenkin, 
{\it Classical and quantum cosmology of the Starobinsky inflationary model},
Phys. Rev. {\bf D32}, 2511
(1985)


\bibitem{nojiri}
S. Nojiri, S.D. Odintsov and V.K. Oikonomou,
 {\it Gravity Theories on a Nutshell: Inflation, Bounce and Late-time Evolution}, 
Phys. Rept. {\bf 692},  1-104 (2017)  [arXiv:1705.11098 [gr-qc]]. 

\bibitem{nojiri1}
S. Nojiri and S. D. Odintsov, 
{\it Unified cosmic history in modified gravity: from $F(R)$ theory to Lorentz non-invariant models}, 
Phys. Rept. {\bf 505}, 59-144 (2011)
 	[arXiv:1011.0544 [gr-qc]].

\bibitem{aho}
J. Amor\'os, J. de Haro and S.D. Odintsov,
{\it On  $R+\alpha R^2$ Loop Quantum Cosmology},
Phys. Rev. {\bf D89}, 104010 (2014) 	[arXiv:1402.3071 [gr-qc]].


\bibitem {guth}
A. Guth,
{\it The inflationary universe: a possible solution to the horizon and flatness problems},  
Phys. Rev. {\bf D 23}, 347 (1981).

\bibitem{Planck}
P.~A.~R.~Ade {\it et al.} 
{\it   Planck 2015 results. XX. Constraints on inflation } 
Astron.\& Astrophys.\  {\bf 594}, A20 (2016)
[arXiv:1502.02114 [astro-ph.CO]].



\bibitem{Starobinsky1}
A. A. Starobinsky, 
Proc. of the Second Seminar {\it Quantum Theory of Gravity} (Moscow, 13-15
Oct. 1981), INR Press, Moscow, 1982, pp. 58-72 (reprinted in: Quantum
Gravity, eds. M. A. Markov, P. C. West, Plenum Publ. Co., New York, 1984,
pp. 103-128)

\bibitem{haro}
J. Haro,
{\it Gravitational particle production: a mathematical treatment}, 
J. Phys A : Math. Theor. {\bf 44}, 205401 (2011).



\bibitem{Gorbunov}
D. S. Gorbunov and  A. G. Panin,
{\it Scalaron the mighty: producing dark matter and baryon asymmetry at reheating}, 
Phys. Lett. {\bf B700}, 157-162 (2011) 
	[arXiv:1009.2448 [hep-ph]]. 

\bibitem{german}
G. German, 
{\it Precise determination of the inflationary epoch and constraints for reheating}, (2020)
 [arXiv:2002.11091 [astro-ph.CO]].











\bibitem{cook}
J. L. Cook, E. Dimastrogiovanni, D. A. Easson and  L. M. Krauss,
{\it Reheating predictions in single field inflation}, 
[arXiv:1502.04673 [astro-ph.CO]].








\bibitem{rg}
T. Rehagen and  G. B. Gelmini,
{\it Low reheating temperatures in monomial and binomial inflationary potentials},
JCAP {\bf 06}, 039 (2015)
	[arXiv:1504.03768 [hep-ph]].


\bibitem{dai}
L. Dai, M. Kamionkowski and J. Wang, 
{\it Reheating Constraints to Inflationary Models},
Phys. Rev. Lett. {\bf 113}, 041302 (2014)
       [arXiv:1404.6704 [astro-ph.CO]].

\bibitem{munoz}
J. B. Mu\~{n}oz and M. Kamionkowski, 
{\it Equation-of-state parameter for reheating},
Phys. Rev. {\bf D91}, 043521 (2015)
       [arXiv:1412.0656 [astro-ph.CO]].


\bibitem{oku}
V. K. Oikonomou, 
{\it Reheating in Constant-roll $F(R)$ Gravity}, 
Mod. Phys. Lett. {\bf A32},  1750172 (2017)
 	[arXiv:1706.00507 [gr-qc]].
	
\bibitem{hashiba}
S. Hashiba and J. Yokoyama, 
{\it Gravitational reheating through conformally coupled superheavy scalar particles}, 
	JCAP {\bf 01}, 028 (2019) 	[arXiv:1809.05410 [gr-qc]].
	
\bibitem{kolb2}	
D. J. H. Chung, E. W. Kolb and A. J. Long,	
{\it Gravitational production of super-Hubble-mass particles: an analytic approach}	
	JHEP {\bf 01}, 189(2019)		[arXiv:1812.00211 [hep-ph]].	


\bibitem{evan}
E. McDonough,
 {\it The Cosmological Heavy Ion Collider: Fast Thermalization after Cosmic Inflation}, 
 (2020)	[arXiv:2001.03633 [hep-th]].
 
\bibitem{Starobinsky2}
A. A. Starobinsky, private communication, May 2020. 


\bibitem{mukhanov}
V. Mukhanov, {\it Physical Foundations of Cosmology}, Cambridge University Press (2005).

\bibitem{btw}
B. A. Bassett, S. Tsujikawa and D. Wands, 
{\it Inflation Dynamics and Reheating},
Rev. Mod. Phys. {\bf 78}, 537 (2006) [arXiv:astro-ph/0507632].


\bibitem{sebastiani}
L. Sebastiani, G. Cognola, R. Myrzakulov, S.D. Odintsov and S. Zerbini,
{\it Nearly Starobinsky inflation from modified gravity}, 
 	Phys. Rev. {\bf D89}, 023518 (2014)  	[arXiv:1311.0744 [gr-qc]].



\bibitem{turner}
 M. S. Turner, 
 {\it Coherent scalar-field oscillations in an expanding universe},
 Phys. Rev. {\bf D28}, 1243 (1983).
 
 




\bibitem{ford}
L. H. Ford, 
{\it Gravitational particle creation and inflation} 
Phys. Rev. {\bf D35}, 2955 (1987).



\bibitem{btw}
B. A. Bassett, S. Tsujikawa and D. Wands, 
{\it Inflation Dynamics and Reheating},
Rev. Mod. Phys. {\bf 78}, 537 (2006) [arXiv:0507632].





\bibitem{riotto}	
A. Kehagias, A. M. Dizgah and  A. Riotto,	
{\it Comments on the Starobinsky Model of Inflation and its Descendants},
Phys. Rev. {\bf  D 89}, 043527 (2014)	
[arXiv:1312.1155 [hep-th]].	
	
	
\bibitem{planck18}	
Y. Akrami et al., 
{\it Planck 2018 results. X. Constraints on inflation}, (2018)
	[arXiv:1807.06211 [astro-ph.CO]].	

\bibitem{roest}
D. Roest,
{\it Universality classes of inflation},
JCAP {\bf 01}, 007 (2014) [arXiv:1309.1285 [hep-th]].







\bibitem{gkr}
G. F. Giudice, E. W. Kolb and  A. Riotto,
{\it Largest temperature of the radiation era and its cosmological implications}, 
 	Phys. Rev. {\bf D 64},   023508 (2001) 	[arXiv:hep-ph/0005123].





\bibitem{husdal}
 L. Husdal,  {\it On Effective Degrees of Freedom in the Early Universe}, Galaxies {\bf 4}, no. 4, 78 (2016)
 	[arXiv:1609.04979 [astro-ph.CO]] .
 
 

\bibitem{pv}
P. J. E. Peebles and  A. Vilenkin, 
{\it Quintessential inflation},
Phys. Rev. {\bf D59}, 063505 (1999)  [arXiv:astro-ph/9810509].

\bibitem{ha}
J. Haro and L. Arest\' e Sal\' o
{\it  The spectrum of Gravitational Waves, their overproduction in quintessential inflation and its influence in the reheating temperature},
(2020)	[arXiv:2004.11843 [gr-qc]].
	


\bibitem{fkl0}
G. Felder, L. Kofman and  A. Linde, 
{\it Instant Preheating}, 
	Phys. Rev. {\bf D59}, 123523 (1999)  	[arXiv:hep-ph/9812289].

\bibitem{fkl}
G. Felder, L. Kofman and  A. Linde, 
{\it    Inflation and Preheating in NO models},
Phys. Rev. {\bf D 60}, 103505 (1999) [arXiv:hep-ph/9903350].

\bibitem{Lazanov}
K. D. Lozanov and  M. A. Amin,
{\it Self-resonance after inflation: oscillons, transients and radiation domination},
Phys. Rev. {\bf D97}, 023533 (2018)  	[arXiv:1710.06851 [astro-ph.CO]].

	
	
	
	
	
	
	
	
	
	







	
	
	














\end{thebibliography}
\end{document}